\documentstyle[twoside,fleqn,espcrc2,epsf]{article}
\newcommand{\AmS}{{\protect\the\textfont2
  A\kern-.1667em\lower.5ex\hbox{M}\kern-.125emS}}

\title{
Large-$q$ expansion 
of the correlation length in the two-dimensional q-state Potts model
}
\author{
H. Arisue
\address{Osaka Prefectural College of Technology, 
26-12 Saiwai-cho, Neyagawa, Osaka 572-8572, Japan}
}
       
\begin{document}

\begin{abstract}
The large-$q$ expansions of the exponential correlation length and 
the second moment correlation length 
for the $q$-state Potts model in two dimensions
are calculated at the first order phase transition point
both in the ordered and disordered phases.
The expansion coefficients in the ordered and disordered phases 
coincide in lower orders for both of the two types of the correlation lengths, 
but they differ a little from each other in higher orders for the second moment
correlation length.
The second largest eigenvalues of the transfer matrix 
have the continuum spectrum both in the ordered and disordered phases 
in the large-$q$ region, which is suggested to be maintained
even in the limit of $q\to 4$ 
from the analysis of the expansion series.

\end{abstract}

\maketitle

\section{INTRODUCTION}

The $q$-state Potts model 
in two dimensions exhibits the first order phase transition for $q>4$. 
The correlation length $\xi_d$ at the phase transition point 
in the disordered phase was solved exactly\cite{Klumper,Buffenoir,Borgs}.
It diverges when $q$ approaches $4$, 
at which the order of the phase transition changes to the second order.
On the other hand, in the ordered phase 
no analytic result is known for the correlation length $\xi_o$ 
at the phase transition point. 
Janke and Kappler\cite{Janke1994} made an conjecture 
from the result of their Monte Carlo simulations 
that $\xi_o=\xi_d$.
Here we make the large-$q$ expansion of the exponential correlation length 
as well as the second moment correlation length at the phase transition point
both in the ordered and disordered phases.

\section{EXPONENTIOAL CORRELATION LENGTH}
We first investigate the exponential correlation length $\xi_o$ 
in the ordered phase.
There is an obstacle to extract the correction to the leading term of the 
large-$q$ expansion 
for the correlation length at the phase transition point 
directly from the correlation function $<{\cal O}(t){\cal O}(0)>_c$, 
since we know from the graphical expansion that it behaves like
\begin{eqnarray}
  <{\cal O}(t){\cal O}(0)>_c & \propto & z^t ( 1 + 2 z t^2 + \cdots ) 
                            \label{eq:graph}\\
                 &\ne& \exp{(- m t)}.\nonumber
\end{eqnarray}
for a large distance $t$ with $z\equiv 1/\sqrt{q}$.
Thus we will diagonalize the transfer matrix $T$ directly for large-$q$.
The eigenfunction for the largest eigenvalue $\Lambda_0$ 
in the leading order is
$$
         |0> \equiv  |\underbrace{0\ 0\ 0\ ..... 0\ 0\ 0}_N>\;
$$
where all of the $N$ spin variables are fixed to be zero,
with the transfer matrix element  $<0|T|0>$ $ = 1 + O(z^2)$
and the corresponding eigenvalue is $\Lambda_0 = 1 + O(z)$.
The eigenfunction for the second largest eigenvalues
are 
$$
      |I> \equiv \frac{1}{\sqrt{N-I+1}}
                \sum |0\ ... 0\ \underbrace{e\ e\ e\ ... e\ e}_{I} 0\ ... 0>
$$
\vspace{-5mm}
$$
               (I=1,\cdots,N),
$$
$$
              |e> \equiv \frac{1}{\sqrt{q-1}} \sum_{s_i=1}^{q-1} |s_i>
$$
with
\begin{eqnarray}
<I|T|I> &=& z + O(z^2), \nonumber\\
<I|T|I+1> &=& 2\sqrt{\frac{N-I}{N-I+1}} z^{3/2}  + O(z^{5/2})\nonumber
\end{eqnarray}
and other matrix elements starting from higher orders. 
We note that the second largest eigenvalues of $T$ degenerate in the 
leading order with $\Lambda_i = z + O(z^{3/2})$.
Diagonalizing $<I|T|J>$ we obtain
$$
 z - 4z^{3/2} \le \Lambda_i \le z + 4z^{3/2} \ (i=1,\cdots,N)
$$
for $N \to \infty$, constituting a continuum spectrum of the eigenvalues. 

    In higher orders, the standard perturbation theory gives
$$\Lambda_1/\Lambda_0 = z + 4 z^{3/2} + 6 z^{2} + O(z^3),$$
from which we obtain
\begin{eqnarray}
1/\xi_o &\equiv & -\log{(\Lambda_1/\Lambda_0)} \nonumber\\
        &=& - \log{z} - 4z^{1/2} + 2z +8/3z^{3/2}+ O(z^2). \nonumber
\end{eqnarray}
This is the same as the large-$q$ expansion of $1/\xi_d$ given 
by Buffenoir and Wallon\cite{Buffenoir} to this order.

    Let us return to the large-$q$ expansion of the correlation function
$$    <{\cal O}(t){\cal O}(0)>_c 
            = \sum_{i=1}^N c_i^2 (\Lambda_i/\Lambda_0)^t\;. $$
with $c_i=<\Lambda_0|{\cal O}(t)|\Lambda_i>$ the overlapping amplitude.
Using  $\Lambda_i/\Lambda_0 = z + f_i z^{3/2} +O(z^2)\ \ \ (i=1,\cdots, N)$
we have
       $$ 
          <{\cal O}(t){\cal O}(0)>_c  
             = \sum_{i=1}^N c_i^2 z^t 
               (1+ f_i z^{1/2}t + f_i^2 z t^2 /2 + \cdots)\;.$$ 
Since $f_i=-f_{N-i+1}$ and  $c_i=c_{N-i+1}$,
the terms proportional to $t$ cancel,
and in the limit of $N\to \infty$  we have
$$    <{\cal O}(t){\cal O}(0)>_c  
             = z^t (1+ 2 z t^2+\cdots)\;,$$ 
which is consistent with the result (\ref{eq:graph}) 
of the graphical expansion. 

 In the disordered phase the situation is quite similar.
The eigenfunction for the largest eigenvalue in the leading order is 
$$
|0> \equiv |\underbrace{e'\ e'\ e'\ ..... e'\ e'\ e'}_N>\;,
$$
\vspace{-2mm}
$$
     \qquad\qquad |e'> \equiv \frac{1}{\sqrt{q}} \sum_{s_i=0}^{q-1} |s_i>\;.
$$
The eigenfunction for the second largest eigenvalues are
\begin{eqnarray}
 |I,f> &\equiv& 
  \sum_i b_i^{(f)} |e'\ ... e'\ \underbrace{i\ i\ i\ ... i\ i}_{I} e'\ ... e'>
        \nonumber\\
       &&   (I=1,\cdots,N ;\quad f=1,\cdots,q-1) \nonumber
\end{eqnarray}
with $\sum_i b_i^{(f_1)} b_i^{(f_2)} = \delta_{f_1,f_2}$.
Diagonalizing \\$<I|T|J>$ we obtain
             $ z - 4z^{3/2} \le \Lambda_i \le z + 4z^{3/2} $
in the limit of $N \to \infty$.
These eigenvalues also constitute a continuum spectrum.

\section{SECOND MOMENT CORRELATION LENGTH}
Here we give the results of the large-$q$ expansion 
of the second moment correlation lengths $\xi_{2nd,o}$ 
in the ordered phase and $\xi_{2nd,d}$ in the disordered phase 
defined by 
$$
\xi_{2nd}^2=\frac{\mu_2}{2d\mu_0}\;,$$
$$\mu_0=\lim_{V\to \infty} \frac{1}{V} \sum_{i,j} <O_i\ O_j>_c\;, $$
$$\mu_2=\lim_{V\to \infty} \frac{1}{V} \sum_{i,j} (i-j)^2<O_i\ O_j>_c\;, $$
$$O_i=\delta_{s_i,0}\;. $$
The finite lattice method 
is used to obtain the large-$q$ expansion.
The algorithm to apply it to the Potts model was 
given by Tabata and H. A.\cite{Arisue1999}
and that for obtaining the second moment correlation length was given 
by Tabata and H. A.\cite{Arisue1995}.
%
%
 The obtained expansion coefficients for the ordered and disordered phases
coincide with each other to order $z^{3}$ 
and differ from  each other in higher orders.
Their ratio estimated by the Pad\'e analysis is given in Fig.1. 
It is not far from unity
even in the limit of $q \to 4$ ($\xi_{2nd,o}/\xi_{2nd,d}=0.930(3)$).
\begin{figure}[tb]
\epsfxsize=7.4cm
\epsffile{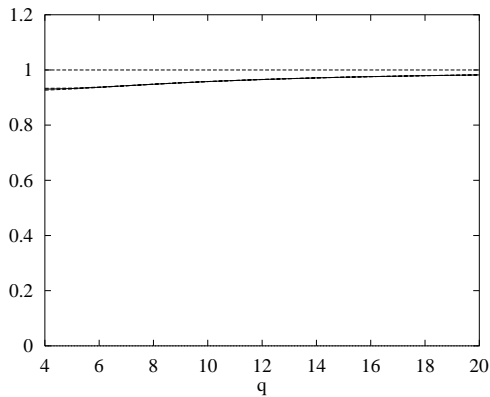}
\vspace{-10mm}
\caption{
Ratio of the second moment correlation lengths 
in the ordered to disordered phases. 
         }
\vspace{-5mm}
\end{figure}
%
%
%
%
%
%
%
%
%
%

\begin{figure}[tb]
\epsfxsize=7.4cm
\epsffile{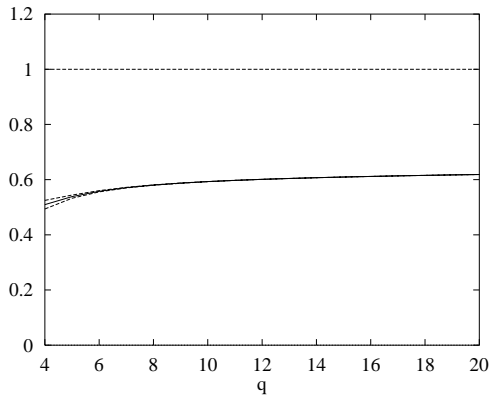}
\vspace{-10mm}
\caption{
Ratio of the second moment correlation length to the 
exponential correlation length 
in the disordered phase. 
         }
\vspace{-5mm}
\end{figure}
Another interesting point is that, as described in Fig.2, 
the ratio $\xi_{2nd,d}/\xi_{d}$ of the second moment correlation length 
to the exponential correlation length 
is much less than unity in the region of $q$ where the correlation length is
large enough. 
It approaches $0.51(2)$ for $q \to 4$. 
It is known that in the limit of the large 
correlation length\cite{Caselle1999}, 
$$
       \frac{\xi_{2nd}^2}{\xi^2}
      \to \frac{\sum_{i=1}^{\infty} c_i^2 (\xi_i/\xi_1)^3}
               {\sum_{i=1}^{\infty} c_i^2 (\xi_i/\xi_1)}
      < 1,
$$ 
with $\xi_i\equiv -\log{(\Lambda_i/\Lambda_0)}$.
If the 'higher excited states' ($i=2,3,\cdots$) did not contribute so much, 
this ratio would be close to unity,
as in the case of the Ising model on the simple cubic lattice,
where $\xi_{2nd}/\xi=0.970(5)$ 
at the critical point\cite{Caselle1999,Campostrini1998}.
Our result implies that the contribution of the 'higher excited states' 
is important in the Potts model in two dimensions.
This is consistent with the fact that the eigenvalue of the 
transfer matrix for the 'first excited state' locates 
at the edge of the continuum spectrum at least in the large-$q$ region.

\section{SUMMARY}

The large-$q$ expansion of the eigenvalues {$\Lambda_i$} 
of the transfer matrix 
was calculated for the two-dimensional $q$ state Potts model 
at the first order phase transition point.
We found that $\xi_o=\xi_d$ at least to order $z^{3/2}$
(i.e. in the first 4 terms).
There are continuum spectra of the eigenvalues of the transfer matrix 
for $N\to \infty$ both in 
the ordered and disordered phases  at least in the large-$q$ region.
We also calculated the large-$q$ expansion of $\xi_{2nd,o}$ and $\xi_{2nd,d}$
and found that $\xi_{2nd,o} \ne \xi_{2nd,d}$ for higher orders than $z^{3}$. 
Note that this does not necessarily imply {$\xi_o \ne \xi_d$},
because the second moment correlation length depends not only on the spectrum
of the eigenvalues of the transfer matrix but also on the overlapping integral.

We also found that $\xi_{2nd,d}/\xi_d $ is far from unity 
for all region of $q>4$. 
It receives significant contributions 
not only from the 'first excited state' but also 'higher excited states', 
which form a continuum spectrum at least in large-$q$ region.
We expect that, even in the limit of $q\to 4$, 
the continuum spectrum would be maintained
(i.e. there would be  no particle state).
In the previous analysis of the Monte Carlo data
it was assumed that the correlation function at long distances 
would be dominated by a single exponentially decaying term.
So they seems necessary to be reanalyzed.

\end{document}